\documentclass[prl,twocolumn,superscriptaddress,showpacs]{revtex4-1}

\usepackage{graphicx}
\usepackage{amssymb, amsfonts, amsmath}
\usepackage{bm}
\usepackage{color}
\usepackage{times}

\begin{document}

\title{Evolution of the Fermi surface of the nematic superconductors FeSe$_{1-x}$S$_x$}

\author{A. I. Coldea}
\email[corresponding author:]{amalia.coldea@physics.ox.ac.uk}
\affiliation{Clarendon Laboratory, Department of Physics,
University of Oxford, Parks Road, Oxford OX1 3PU, UK}

\author{S.\,F.\;Blake}
\affiliation{Clarendon Laboratory, Department of Physics,
University of Oxford, Parks Road, Oxford OX1 3PU, UK}

\author{S.\;Kasahara}
\affiliation{Department of Physics, Kyoto University, Kyoto 606-8502, Japan}

\author{A. A.\;Haghighirad}
\affiliation{Clarendon Laboratory, Department of Physics,
University of Oxford, Parks Road, Oxford OX1 3PU, UK}

\author{M. D.\;Watson}
\affiliation{Clarendon Laboratory, Department of Physics,
University of Oxford, Parks Road, Oxford OX1 3PU, UK}
\affiliation{Diamond Light Source, Harwell Campus, Didcot, OX11 0DE, UK}

\author{W.\;Knafo}
\affiliation{Laboratoire National des Champs Magn\'{e}tiques Intenses (LNCMI-EMFL), UPR 3228, CNRS-UJF-UPS-INSA,
143 Avenue de Rangueil, 31400 Toulouse, France}

\author{E. S.\; Choi}
\affiliation{National High Magnetic Field Laboratory, Florida State University, Tallahassee, FL 32310, USA}

\author{A.\;McCollam}
\affiliation{High Field Magnet Laboratory (HFML-EMFL), Radboud University, 6525 ED Nijmegen, The Netherlands}

\author{P. Reiss}
\affiliation{Clarendon Laboratory, Department of Physics,
University of Oxford, Parks Road, Oxford OX1 3PU, UK}

\author{T.\;Yamashita}
\affiliation{Department of Physics, Kyoto University, Kyoto 606-8502, Japan}

\author{M.\;Bruma}
\affiliation{Clarendon Laboratory, Department of Physics,
University of Oxford, Parks Road, Oxford OX1 3PU, UK}

\author{S.\;Speller}
\affiliation{Department of Materials, University of Oxford, OX1 3PH, UK}

\author{Y.\;Matsuda}
\affiliation{Department of Physics, Kyoto University, Kyoto 606-8502, Japan}

\author{T.\;Wolf}
\affiliation{Institut f\"{u}r Festk\"{o}rperphysik, Karlsruhe Institute of Technology, 76021 Karlsruhe, Germany}

\author{T.\;Shibauchi}
\affiliation{Department of Advanced Materials Science, University of Tokyo, Kashiwa, Chiba 277-8561, Japan}

\author{A. J. Schofield }
\affiliation{School of Physics and Astronomy, University of Birmingham, Edgbaston, Birmingham B15 2TT, UK}

\begin{abstract}

We investigate the evolution of the Fermi surfaces and electronic interactions across the nematic phase transition in single crystals of FeSe$_{1-x}$S$_x$ using Shubnikov-de Haas oscillations in high magnetic fields up to 45 tesla  in the low temperature regime. The unusually small and strongly elongated Fermi surface of FeSe increases monotonically with chemical pressure, $x$, due to the suppression of the in-plane anisotropy except for
the smallest orbit which suffers a Lifshitz-like transition once nematicity disappears.
Even outside the nematic phase the Fermi surface continues to increase, in stark contrast to the reconstructed Fermi surface detected in FeSe under applied external pressure. We detect signatures of orbital-dependent quasiparticle mass  renomalization suppressed for those orbits with dominant $d_{xz/yz}$ character, but unusually enhanced for those orbits with dominant $d_{xy}$  character.
 The lack of enhanced  superconductivity outside the nematic phase in FeSe$_{1-x}$S$_x$ suggest that nematicity may not play the essential role in enhancing $T_c$ in these systems.
\end{abstract}

\date{\today}

\maketitle

\begin{figure*}[t]
	\includegraphics[width=1\linewidth]{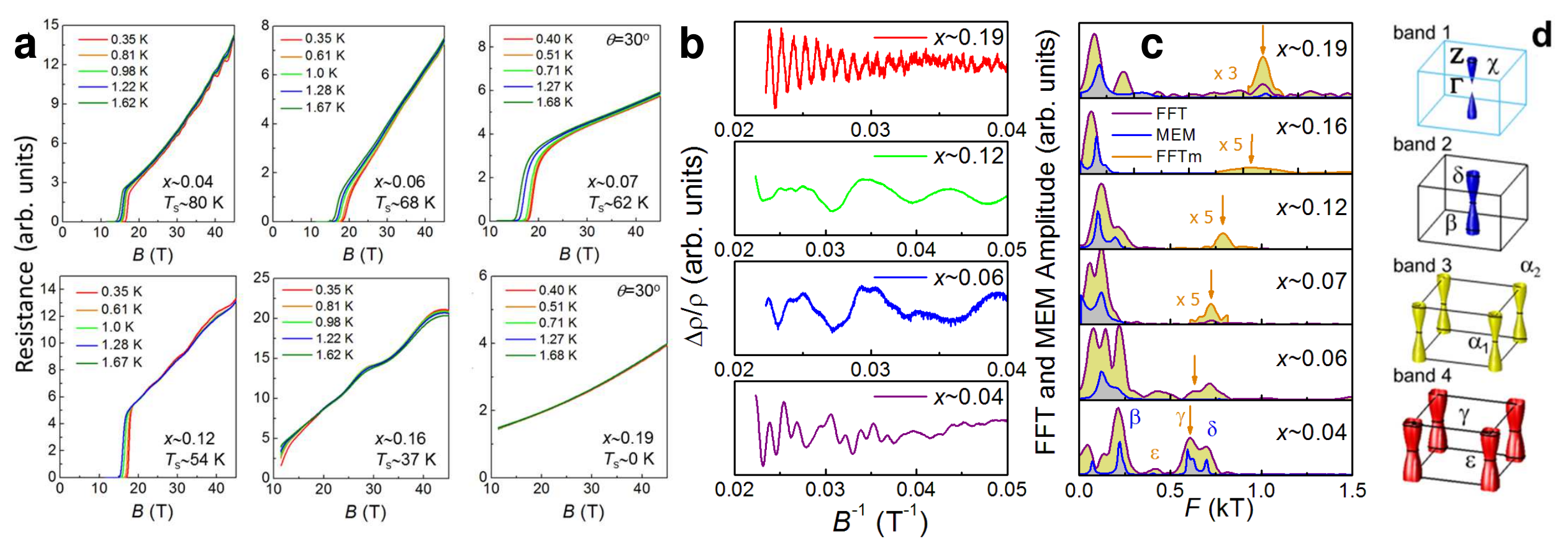}
	\caption{{\bf Quantum oscillations in superconducting single crystals of FeSe$_{1-x}$S$_x$.} {\bf a},  The in-plane resistance $R_{xx}(B)$ as a function of magnetic field $B$ for different compositions, $x$ and $T_s$.  Measurements were performed at constant low temperatures between 0.3-2~K by sweeping the magnetic fields up to 45~T with  $B || c$ axis ($\theta$=0) or $\theta=30^{\circ}$ for $x$=0.19 and $x$=0.07.
b, The oscillatory part of the resistivity $\Delta \rho_{osc}$/$\rho$ obtained by removing a polynomial background from the raw data in a) at the lowest measured temperature. c, The frequency spectra obtained using a fast Fourier transform (FFT) and maximum entropy method (MEM). A multiplied FFT spectrum is used to emphasize the high frequency features with their positions indicated by vertical arrows. d, The proposed Fermi surface and the different extremal areas for FeSe$_{1-x}$S$_x$ obtained by shrinking the calculated Fermi surface in the tetragonal phase by more than 150 meV \cite{Watson2014,Watson2015} (see also Supplemental Material). Frequencies below 200~T cannot be reliably assigned due to a possible overlap of at least 3 different small frequencies ($\alpha_1$, $\alpha_2$ and $\chi$).}
	\label{fig1}
\end{figure*}

Nematic electronic order is believed to
play an important role in understanding superconductivity in iron-based and copper oxide superconductors \cite{Fernandes2013,Schattner2016,Wang2015c}.
FeSe is an ideal system to study this
interplay as the nematic phase below 87~K is manifested via a strong distortion of the Fermi surface, which has been linked to the lifting of the orbital degeneracy of the Fe $d_{xz}$/$d_{yz}$ bands \cite{Watson2014} and the development of $d_{xy}$ orbital anisotropy \cite{Watson2016}.
Another peculiarity of FeSe is the strong disparity in band renormalization
between the degenerate  $d_{xz/yz}$ bands compared with  a much stronger effect
 for the $d_{xy}$ band,  as determined from angle resolved photoemission spectroscopy (ARPES) at high temperatures \cite{Watson2014}.
 In bulk FeSe, superconductivity emerges out of the electronic nematic phase  \cite{Watson2014}, but its relevance for the high $T_c$ superconductivity found under applied pressure, intercalation or by surface doping, remains unclear \cite{Wen2016,Terashima2016,Clarke2013}.
One approach in finding the interplay between nematicity and superconductivity is using
 chemical pressure induced by isoelectronic substitution in FeSe$_{1-x}$S$_x$, which
 dramatically suppresses the nematic phase  \cite{Watson2015}.

  Chemical pressure may mimic the behaviour of applied hydrostatic
pressure by bringing the FeSe layers closer together, potentially
increasing the bandwidth and suppressing the electronic correlations.
Thus, isoelectronic substitution in superconducting FeSe$_{1-x}$S$_x$
is an alternative tuning parameter which may provide new insight into the interplay
of the nematic order found in the parent compound, and the induced
magnetic order and enhanced superconductivity found under physical pressure
\cite{Terashima2015,Knoner2015,Kothapalli2016}.
In this paper we use quantum oscillations in a series of sulphur-substituted FeSe single crystals to
probe in detail the evolution of complex Fermi surfaces and
quasiparticle properties across the putative nematic quantum phase
transition at very low temperatures and high magnetic fields.
Samples were grown by the KCl/AlCl$_3$ chemical vapour transport method.
and their compositions were determined using EDX measurements.
Magnetotransport measurements  were performed
in high magnetic fields up to 45~T at NHMFL in Tallahassee and up to 33~T at HFML in Nijmegen
at low temperatures down to 0.35~K using $^3$He cryogenic systems.


{\it Quantum Oscillations in FeSe$_{1-x}$S$_x$.}
Fig.~\ref{fig1} shows resistivity measurements of FeSe$_{1-x}$S$_x$
in magnetic fields up to 45~T and at low temperatures ($T \sim 0.4$~K),
with the magnetic field either perpendicular to the highly conducting $ab$ layers ($B\||c$, $\theta$=0) or tilted away by an angle ($\theta$=$30^{\circ}$).
The upper critical field for the transition from the superconducting to the normal state in FeSe$_{1-x}$S$_x$ reaches
a maximum of about 20~T ($x \sim 0.07(1)$ for $B\||c$) and it varies with composition in a similar way as $T_c$, which has a maximum
of about 11(1)~K for the same composition, as shown in the phase diagram in Fig.\ref{fig3}a.
In the normal state, we observe Shubnikov-de Haas oscillations
superimposed on the transverse magnetoresistance signal,
with different samples showing dominance of different
frequencies due to the multiple electronic bands involved across the compositional range.
The quantum oscillations are periodic in $1/B$ and this is clearly visualized
after removing a polynomial background magnetoresistance, as shown in Fig.\ref{fig1}b.
Quantum oscillations are a bulk probe of the size and shape of the Fermi surface,
by measuring extremal cross-sectional orbits with a high accuracy of up to $\sim$ 0.04 \% of the area of the first Brillouin zone. The extremal areas of the Fermi surface normal to the applied magnetic field are obtained
from the Onsager relation, $F_i = A_{ki} \hbar /(2 \pi e)$, as each pocket can contribute two
extremal orbits to the quantum oscillation signal
(minimum at $k_z$=0 and maximum at $k_z$=$\pi/c$) \cite{Shoenberg}.

The low-temperature experimental Fermi surface of FeSe is that of a multi-band system, being composed of small and anisotropically distorted quasi-two dimensional electronic bands, one hole and two electron bands \cite{Watson2014}, with the smallest electron bands showing unusually high mobility \cite{Watson2015PRL,Huynh2014}.
The fast Fourier transform (FFT) spectra of such a complex Fermi surface could have up to six different frequencies, with two different frequencies, respectively, associated with the minimum and maximum extremal orbits of each quasi-two dimensional cylindrical Fermi surface.
Fig.\ref{fig1}c shows the FFT spectrum at low concentrations ($x \sim$ 0.04) for FeSe$_{1-x}$S$_x$, which is quite similar to
previous studies on FeSe \cite{Terashima2014,Watson2014} displaying many frequencies, labelled as $\alpha$, $\beta$, $\gamma$ and $\delta$, $\epsilon$, all below 800~T. This suggests
that each pocket of the Fermi surface corresponds
 to a very small fraction of $\sim 2\% $ of the basal plane area of the Brillouin zone (see Fig.\ref{fig1}d)
 and these pockets areas are a factor $\sim 5$ smaller than those predicted by the band structure calculations (see  Fig.SM7 in Supplemental Material (SM)).
 The observed frequencies for $x \sim 0.04$ are associated with the corrugated hole-like cylinder, with extremal areas $\beta$ (214~T) and $\delta$ (690~T) and two electron-like pockets
 that have very different quasiparticle effective masses.
  The outer electron cylindrical band with dominant $d_{xy}$ character
 has a maximum extremal orbit, $\gamma$ ($\sim$609~T), and a possible minimum orbit $\epsilon$ ($\sim $ 440~T)
which has a similar mass of $\sim6 m_e$ to $\gamma$, and it is unlikely to be
a second harmonic of $\beta$ (see Fig. SM4).
 The inner small electron band, pocket $\alpha$, should also be a cylinder with two possible orbits $\alpha_1$  and $\alpha_2$, as drawn in Fig.\ref{fig1}d).
However, these low frequencies are  often difficult to distinguish in the FFT spectra of FeSe and their signatures 
  can be seen below 200~T in Fig.\ref{fig1}c, where both the polynomial background as well as $1/f$ noise
  can generate artificial peaks.  This small $\alpha$ pocket is expected
  to have a light mass and high mobility \cite{Watson2015PRL}.

\begin{figure}[ht]
	\centering
\includegraphics[width=1\linewidth,clip=true]{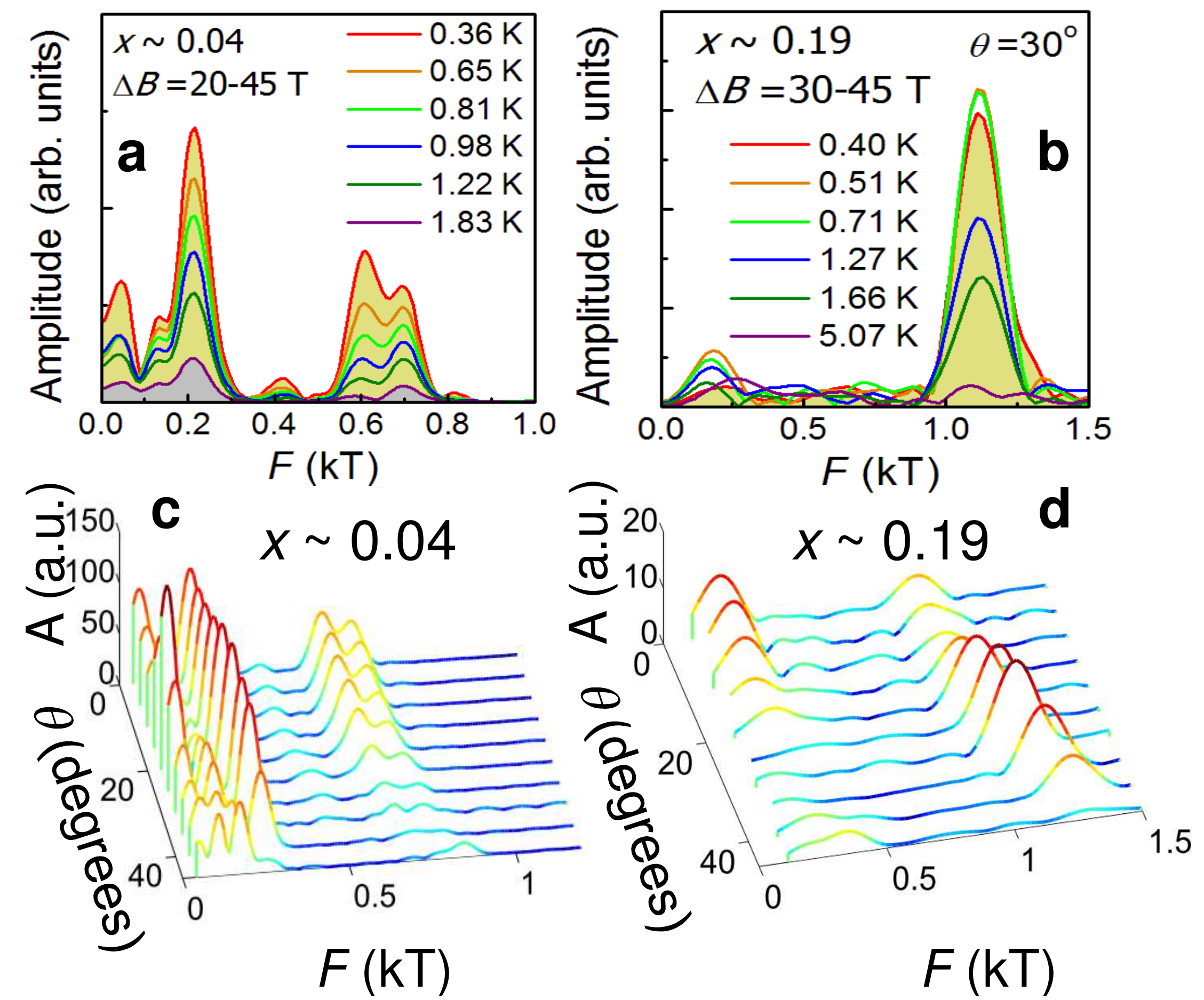}
\includegraphics[width=1\linewidth,clip=true]{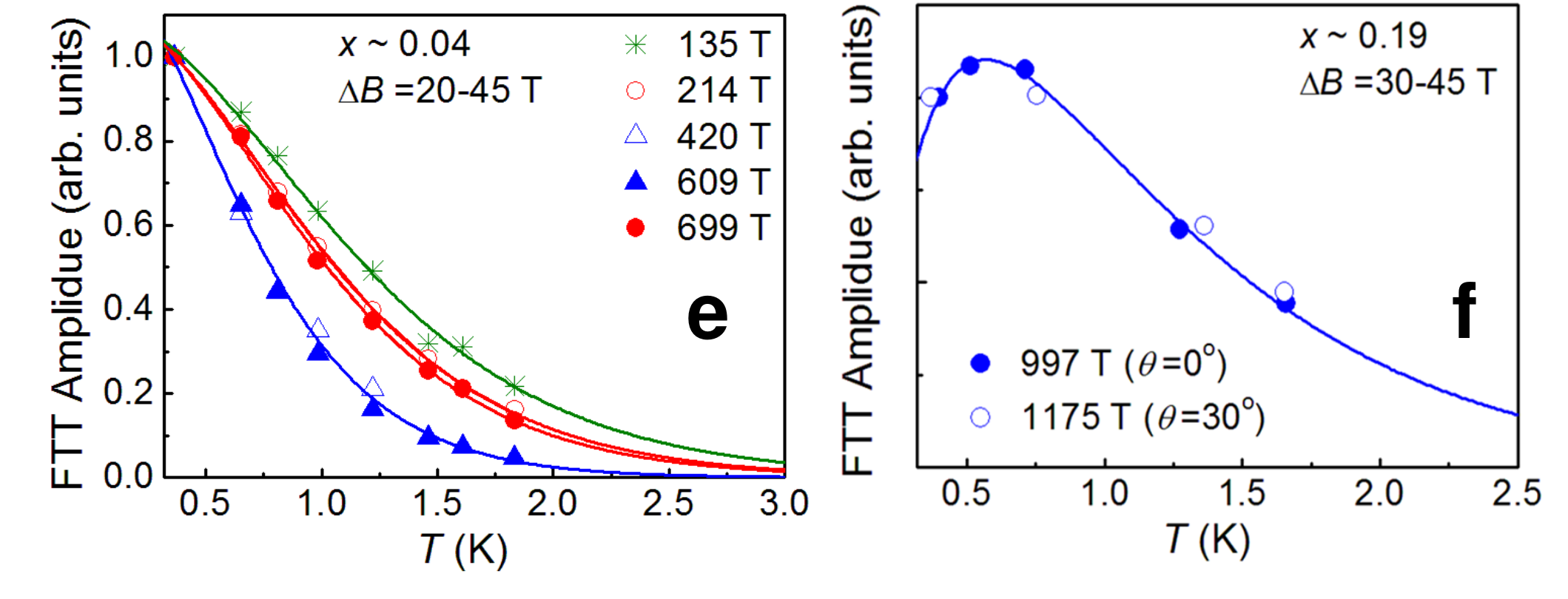}
	\caption{\textbf{Quasiparticle effective masses of FeSe$_{1-x}$S$_x$.} a, b Temperature-dependence of the Fourier transform (FFT) amplitude of different frequencies for two different compositions, $x \sim$ 0.04 with $T_s \sim$ 80~K and $x \sim$ 0.19, which is outside the nematic phase. c,d The angular dependence of the FFT amplitudes showing for the highest frequency
an enhanced amplitude for $x \sim$ 0.19  around $\theta \sim 30^{\circ}$. e, f, Temperature-dependence of the amplitude of oscillation from which the effective masses is extracted using a Lifshitz-Kosevich (LK) formula \cite{Lifshitz1956} (solid line) for different compositions. Clear deviation from the LK formula is observed for $x \sim$ 0.19 and the fit uses a two-component model, as described in Ref.\cite{McCollam2005}.}
	\label{fig2}
\end{figure}

With increased sulphur substitution around $x \geq 0.12$, the observed magnetoresistance is
indeed dominated by a low frequency, which decreases from 120(5)~T for $x \sim 0.12$ to 67(5)~T for $x \sim 0.16$
and around 35(5) for $x \sim 0.17$, as
seen in Fig.\ref{fig1}a and Fig.SM3.
An unusual feature of this small pocket for $x \sim$0.16 is its
non-trivial Berry phase and a lighter mass ($\sim$ 1.7(1) $m_e$), which may originate
from a Dirac-like dispersion of a small electronic band
($\alpha_1$ or $\alpha_2$ in Fig.\ref{fig1}d) which
would disappear outside the nematic phase \cite{Kasahara2016}.
On the other hand, this prominent low frequency could also originate from the inner hole
band centered at the $Z$ point ($\chi$ in Fig.\ref{fig1}d), which was detected in ARPES studies for $x \sim 0.12$,
once the orbital ordering is reduced by the sulphur substitution
\cite{Watson2015}.
The high frequencies of the outer electron and hole bands
($\gamma$ and $\delta$, respectively)  increase monotonically with $x$, even outside the nematic phase
as shown in Fig.~\ref{fig1}c, implying that there is no Fermi surface reconstruction
induced by chemical pressure due to the lack
of a magnetic phase outside the nematic phase,
in contrast to the studies on FeSe under applied pressure \cite{Terashima2016}.
Instead, our quantum oscillation data for FeSe$_{1-x}$S$_x$
may suggest that a Lifshitz-like transition may take place
as we observe the disappearance of a low frequency
once the nematic phase is suppressed (see also Fig.\ref{fig3}). This effect  occurs 
when a cylindrical Fermi surface
topology changes due to a strong increase of the interlayer warping. Consequently, 
Fermi surface becomes disconnected in the centre of the Brillouin zone
and one candidate for this behaviour would be $\alpha_1$ in Fig.\ref{fig1}d.
The presence of a Lifshitz-like transition on the edge of the nematic phase
was previously identified in ARPES studies in electron-doped single crystals of FeSe
using in-situ potassium coating \cite{Ye2015}. A Lifshitz transition
was found to coincide with the disappearance of
the spin-density wave in electron-doped BaFe$_2$As$_2$ \cite{Liu2010}.

{\it Quasiparticle effective masses.}
A direct manifestation of the effect of electron-electron interactions
is the relative enhancement of the quasi-particle effective masses compared to the bare band mass.
The cyclotron-averaged effective masses of the quasiparticles
for each extremal orbit can be extracted from the temperature dependence of the 
amplitude of the quantum oscillations \cite{Shoenberg},
as shown in Fig.\ref{fig2} and Fig.SM4.
For low substitution $x\sim0.04$, the values of the effective masses
are close to those reported for FeSe \cite{Terashima2014,Audouard2014}; the $\beta$ and $\delta$ orbits of the outer hole band have similar effective masses $\sim 4(1)~m_e$, indicating that they originate from the same band with dominant $d_{xz}/d_{xz}$ band character at the minimum and maximum $k_z$ locations of a corrugated quasi-two dimensional band. The
   $\gamma$ and $\epsilon$ orbits correspond to the outer electron band with dominant $d_{xy}$ band character, which has a particularly heavy mass of $\sim 7(1)$ $m_e$, confirming the strong
orbitally-dependent band renormalization found by ARPES studies \cite{Watson2014}.

Outside the nematic phase for $x\sim0.19$,  the background magnetoresistance
is almost quadratic and we detect some frequencies below 600~T (see also Sample 2 in Fig.SM6), but
 only a single high frequency oscillation with a rather peculiar behaviour, as shown in Fig.\ref{fig2}b and d.
Firstly, the largest amplitude of this frequency is observed not at $\theta=0$, as expected for a Shubnikov-de Haas
signal but at $\theta=30^{\circ}$ (Fig.\ref{fig2}d). This angle is not the Yamaji angle, ($\theta_Y=70^{\circ}$) \cite{Yamaji1989}
at which a constructive interference occurs when the minimum and maximum orbit of a quasi-two dimensional cylinder coincide
\footnote{The Yamaji angle caused by the interference of the maximum and minima orbits of a corrugated Fermi surface occurs when $J_0$($c k_F \tan \theta_Y) = 0$, where $J_0$ is the Bessel function,  $k_F$ is the in-plane Fermi momentum and $c$ is the lattice parameter.}.
 Other reasons for an interference could be that coincidentally two orbits, originating either from the outer
  hole and electron bands, generate the same frequencies (and cannot be separated by the FFT due to limited field window) or that both frequencies originate from the same band, but are spin-split in the presence of magnetic order, as the Zeeman energy alone would not generally generate two frequencies (see also the splitting shown by the MEM spectrum in Fig.SM6).
Secondly, the temperature dependence of the FFT amplitude of this high frequency (Fig.\ref{fig2}e) drops at low temperatures (below 0.6~K), in contrast to the expected Lifshitz-Kosevich behaviour found for all frequencies
for $x \sim $ 0.04 (Fig.\ref{fig2}f). This behaviour is reminiscent of the low temperature behaviour of heavy fermion systems with strongly polarized bands and close frequencies, but different effective masses \cite{McCollam2005}. Thus, to
model the observed temperature dependence we use an extended Lifshitz-Kosevich formula to account
  for the phase difference of the two orbits \cite{McCollam2005}, which gives
 two very different cyclotron effective mass: one lighter mass of $\sim 3.7 m_e$ (dominant at high-temperature) likely assigned to the outer hole band with $d_{xz}/d_{yz}$  character and a much heavier mass, possibly larger than $\sim 15 (7) m_e$
 (see the deviation of the FFT amplitude below 0.5~K in Fig.SM4). This suggest a strong mass enhancement for the outer electron band with dominant $d_{xy}$ band character just outside the nematic phase.  An alternative interpretation could be that we detect a spin-split Fermi surface in the presence of a local effective magnetic field, as in heavy fermions \cite{McCollam2005}; but as there is no magnetism detected in these materials outside the nematic phase, this scenario may be unlikely.
 The effective masses and electronic correlations associated to the orbits with predominant $d_{xz}/d_{yz}$  character (outer hole band in Fig.\ref{fig3}d)
 are suppressed  with $x$ (with  no clear signatures of criticality),  similarly to the  evolution of $T_c$ (Fig.\ref{fig3}a), suggesting
 that these bands play an dominant in the pairing mechanism and their strength is reduced
 once the electronic bandwidth increases with chemical pressure \cite{Reiss2017}.

\begin{figure}[ht]
	\centering
\includegraphics[trim={1cm 1cm 11cm 1cm},width=1\linewidth,clip=true]{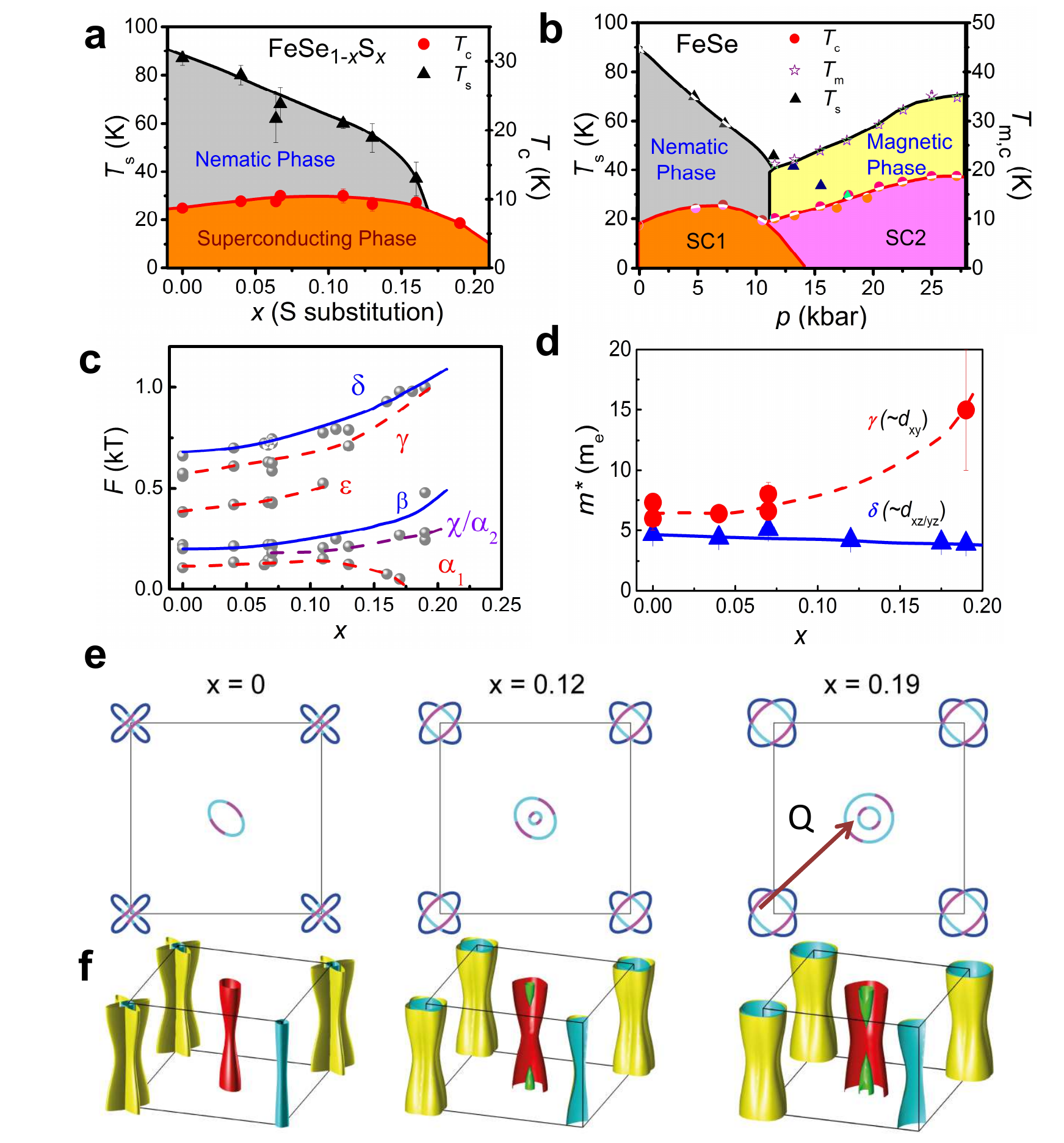}
	\caption{\textbf{Evolution of the electronic properties of FeSe$_{1-x}$S$_x$.} a, Phase diagram of FeSe$_{1-x}$S$_x$ indicating the suppression of the nematic phase, ($T_s$, solid triangles) and the small dome of superconductivity induced by sulphur substitution, $x$ ($T_c$, solid circles) using data from Fig.SM1.
b, Proposed phase diagram of FeSe under pressure based on Ref. \cite{Terashima2016} showing
the low $T_c$ (SC1) and high $T_c$ phases (SC2) as well as the new magnetic phase.
c, The evolution of the observed quantum oscillations frequencies. Solid lines are the calculated
frequencies for the large outer hole band based on ARPES data from Ref.\cite{Watson2015}. The dashed lines are guides to the eye.
d, The quasiparticle effective masses of the high frequencies, $\gamma$ and $\delta$. Solid and dashed lines are guides to the eye. e and f, Proposed Fermi surfaces of FeSe$_{1-x}$S$_x$ for different values of $x$ as a slice at $k_z=\pi/c$ and 3D representation, respectively. A possible inter-band nesting vectors, $\bm Q$ is also shown.}
	\label{fig3}
\end{figure}

{\it Fermi surface evolution.}
The  evolution of the low temperature Fermi surfaces
and quasiparticle masses with chemical pressure in FeSe$_{1-x}$S$_x$, summarized in Fig.\ref{fig3},
suggesting a change from a three-band model for FeSe to a four-band model around $x=0.12$, as found in ARPES studies \cite{Watson2015}.
Furthermore, once the boundaries of the nematic phase have been crossed around $x \sim $ 0.18(1), the Fermi surface
suffers a Lifshitz-like transition (as the lowest frequency disappears) whereas the high frequencies corresponding to the outer electron, $\gamma$, and hole, $\delta$, maximum orbits increase monotonically with
chemical pressure, even away from the nematic phase boundaries, as shown in Fig.\ref{fig2}b
and Fig.\ref{fig3}c. This is in contrast to
the Fermi surface of FeSe under applied pressure \cite{Terashima2016},
where only small frequencies were detected outside the nematic phase and were
linked to a possible Fermi surface reconstruction in the presence of the new structural and magnetic phase \cite{Knoner2015,Kothapalli2016}.
With chemical pressure, the Fermi surface sheet of the outer hole band which is severely
elongated for FeSe (see Fig.\ref{fig3}e and f) becomes more isotropic. As the ellipse transform into a circle,
the cross-section area of the Fermi surface increases,
in agreement with the changes in the $k_{\rm F}$ values
determined from ARPES studies \cite{Watson2015} (see solid lines in Fig.\ref{fig3}c).
Additionally to the effect of nematic order on the Fermi surface,
we observe that the orbits corresponding to the outer bands
would continually expand with increasing $x$, and
eventually could reach the calculated values of the same orbits of the superconductor FeS,
as shown by a linear extrapolation for $\gamma$ in Fig.SM7.
 The Fermi surfaces of FeSe$_{1-x}$S$_x$ being severely reduced in size compared
  with the prediction of band structure calculations
  (varying from a factor of five for FeSe towards
 a factor 3 around $x \sim 0.17$)
  is an important consequence of strong orbitally-dependent inter- and intra-band electronic interactions, significantly large in iron chalcogenides \cite{Fanfarillo2016,Watson2014}, but also found in many iron-based superconductors \cite{Shishido2010,Coldea2008}.
  These effects may be suppressed once the bandwidth increases with sulphur substitution from FeSe towards FeS.
 The orbital-dependent mass renormalizations,
 in which the quasiparticle mass of the outer electron band with dominant $d_{xy}$ orbital character ($\gamma$ pocket) seems
to increase with chemical pressure outside the nematic phase, whereas that of the outer hole band with mixed $d_{xz/yz}$ character ($\delta$ pocket) decreases (Fig.\ref{fig3}b),  are likely to play a different role
 in pairing interactions in these relatively low-$T_c$ superconductors as well
 as in enhancing quantum nematic fluctuations, close
  to a putative nematic critical point \cite{Hosoi2016,Wang2016}.
Furthermore, one can envisage that
orbital-dependent effects would promote the stabilization of other competing electronic phases, as found in FeSe under pressure,
but not yet revealed in FeSe$_{1-x}$S$_x$ up to the studied compositions.

FeSe, with a very small Fermi surface, harbours a fragile balance of competing interactions, which either
distort the Fermi surface as a Pomerachuk instability towards a nematic phase, or stabilize a spin-density wave
and/or superconducting phase \cite{Chubukov2016,Mukherjee2015,Yamakawa2016}.
The spin fluctuations increase only at low temperature in the nematic phase of FeSe \cite{Baek2015},
but as the sizes of the outer electron and hole bands coincide outside the nematic phase in
FeSe$_{1-x}$S$_x$, one may expect nesting and spin fluctuations may be enhanced.
The fact that superconductivity is continuously suppressed with chemical pressure 
(Fig.\ref{fig3}a)
could imply that the nematic fluctuations are not the main
factor in superconducting pairing in FeSe$_{1-x}$S$_x$.
The low $T_c$ superconductivity region of FeSe$_{1-x}$S$_x$ has the same electronic
trends as those found in FeSe under pressure below 10~kbar (Fig.\ref{fig3}b), suggesting
that the high $T_c$ superconductivity observed at high pressure
may be decoupled from the nematic state of FeSe by a new structural and magnetic phase \cite{Kothapalli2016}.
This new clean system, FeSe$_{1-x}$S$_x$, provides access to a nematic quantum phase
transition, quite unique for iron-based superconductors, and it
may help clarify the role of nematicity in relation to superconductivity and
the relevance of the Lifshitz
transition and orbitally-selective electronic correlations.

{\it Acknowledgments.}  We thank J. C. A. Prentice for computational
support,  N. Davies and A. Narayanan for preliminary sample preparation
and A. Chubukov and A. Shekhter for useful discussions.
This work was mainly supported by EPSRC (EP/L001772/1, EP/I004475/1, EP/I017836/1).
A.A.H. acknowledges the financial support of the Oxford Quantum
Materials Platform Grant (EP/M020517/1).
A portion of this work was performed at the National High Magnetic Field Laboratory, which is supported by National Science Foundation Cooperative Agreement No. DMR-1157490 and the State of Florida.
Part of this work was supported supported by
HFML-RU/FOM and LNCMI-CNRS, members of the European Magnetic Field
Laboratory (EMFL) and by EPSRC (UK) via its membership to the EMFL
(grant no. EP/N01085X/1). The authors would like to acknowledge
the use of the University of Oxford Advanced Research Computing (ARC)
facility in carrying out part of this work. AIC acknowledges an EPSRC Career Acceleration Fellowship (EP/I004475/1).

\bibliography{FeSe_bib_nov2016}

\end{document}